\documentclass[aps,twocolumn,showpacs,floatfix]{revtex4}

\usepackage{amsmath,amsfonts,amssymb}
\usepackage{graphicx}

\usepackage{color}
\usepackage[normalem]{ulem}
\def\0#1{\xout{#1}}
\def\1#1{{\color{red}#1}}
\def\2#1{{\color{blue}#1}}
\def\3#1{{\color{magenta}#1}}
\def\4#1{{\color{yellow}#1}}
\begin{document}

\title{On the predictability of rogue events}

\author{Simon Birkholz}
\affiliation{Max-Born-Institut, Max-Born-Stra\ss e 2A, 12489 Berlin, Germany}

\author{Carsten Br\'ee}
\affiliation{Weierstra\ss-Institut, Mohrenstr.~39, 10117 Berlin, Germany}

\author{Ayhan Demircan}
\affiliation{Leibniz-Universit\"at Hannover, Welfengarten 1, 30167 Hannover, Germany}

\author{G\"unter Steinmeyer}
\affiliation{Max-Born-Institut,  Max-Born-Stra\ss e\,2A, 12489 Berlin, Germany}
\date{\today}

\begin{abstract}
\noindent Using experimental data from three different rogue wave supporting systems, determinism and predictability of the underlying dynamics are evaluated with methods of nonlinear time series analysis. We included original records from the Draupner platform in the North Sea as well as time series from two optical systems in our analysis. One of the latter was measured in the infrared tail of optical fiber supercontinua, the other in the fluence profiles of multifilaments. All three data sets exhibit extreme-value statistics and exceed the significant wave height in the respective system by a factor larger than two. Nonlinear time series analysis indicates a different degree of determinism in the systems. The optical fiber scenario is found to be driven by quantum noise whereas rogue waves emerge as a consequence of turbulence in the others. With the large number of rogue events observed in the multifilament system, we can systematically explore the predictability of such events in a turbulent system. We observe that rogue events do not necessarily appear without a warning, but are often preceded by a short phase of relative order. This surprising finding sheds some new light on the fascinating phenomenon of rogue waves.
\end{abstract}


\maketitle

\noindent Ocean rogue waves are rare events with extreme magnitude and can pose a serious threat to large ships in the ocean. Their existence has been questioned for centuries until one such event was actually observed on the Draupner platform in the North Sea on January 1, 1995 \cite{Draupner,Draupner2}. About ten years later, a similar statistical anomaly was observed in optical fiber supercontinua \cite{Solli}. Subsequently, the topic of rogue waves received rapidly growing attention, and heavy-tailed probability density functions characteristic for rogue waves have been reported in a multitude of completely different physical systems. There appears to be consent that one qualifying criterion for rogue wave statistics is the appearance of large-amplitude events at far higher frequency than expected by Gaussian statistics \cite{discussion}. In turn, probability density functions exhibit a heavy tail, with a non-vanishing probability of events with extreme magnitude. A second frequently stated property of rogue waves is their surprising appearance and disappearance. Rogue waves have been claimed to ``appear from nowhere and disappear without a trace'' \cite{Akhmediev}. This pictorial statement touches the question whether, at least in principle, a prediction of rogue waves is possible \cite{predict,predict2}. This discussion has been strongly inspired by certain non-stationary solutions of the underlying wave equations, namely Akhmediev breathers and the Peregrine soliton \cite{Peregrine,EarlyWarning,Chabchoub}, which deterministically evolve from a low-level waveform to a transient spike with high peak amplitude. Here we tackle the problem of predictability from a different perspective.  We employ nonlinear time-series analysis \cite{Grassberger,NLtimeseries} to directly gauge the predictability in experimental data. To this end, we use a proven method for the detection of even faint traces of determinism \cite{Grassberger}. This method enables a quantification of rogue wave predictability for the first time.

We base our analysis on experimental data from three different sources. All data sets exhibit a heavytail statistical distribution of events. The first time series has been recorded in the spatial profile of a laser beam that underwent multiple filamentation in a gas cell \cite{OurPRL}. The underlying dynamics is caused by turbulence inside the gas, which is induced by local heating due to the plasma strings inside the multifilament \cite{Milchberg,Dantus}. Time series of the spatial fluence distribution were measured with a camera in the far field of the multiple filaments. Histogram analysis of individual camera pixels indicates a sub-exponential heavy tail, and common qualifiers for rogue wave behavior are exceeded by a large number. In this system, we can find events that exceed the significant wave height (SWH) by a factor up to ten. The SWH is defined as the average of the largest third of waves in a record, and surface water waves are usually considered rogue when they exceed the SWH by a factor of $\approx 2$. The filament data was discretely sampled at an acquisition rate of 1\,kHz synchronized to the laser repetition rate. The data exhibits a linear correlation time constant of about 10\,ms, which corresponds to ten laser shots. This non-zero correlation time sets the filamentation data apart from the behavior of the second system, i.e., real-time spectra that have been recorded at the output of a fiber supercontinuum \cite{Solli2}. The latter data has been recorded at the much higher megahertz repetition rate of a mode-locked oscillator. These spectra exhibit a number of rogue events in their infrared tail. We find events that exceed the SWH by a factor of 4, which clearly qualifies the observed dynamics in the fiber as rogue. Finally, the third data set constitutes of original surface elevation time series recorded in January 1995 on the Draupner oil platform \cite{Draupner,Draupner2}. These relatively short data sets contains rogue events that exceed the SWH by a factor of 2.15 to 2.3.

\begin{figure}[tb]
\centering
\includegraphics[width=\linewidth]{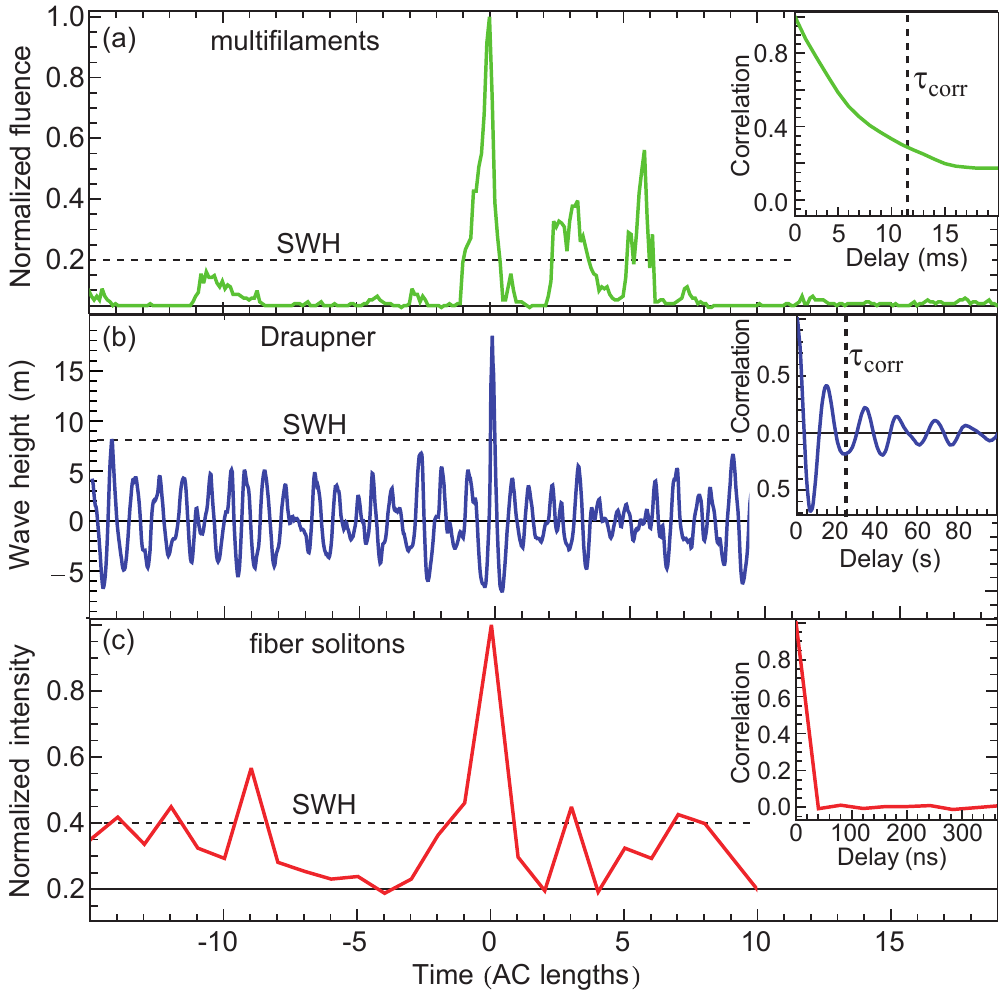}
\vspace{-1em}
\footnotesize{\caption{Examples of rogue events (a) in multifilament dynamics \cite{OurPRL}, (b) the oceanographic context (Draupner1 data set \cite{Draupner,Draupner2}), and (c) soliton dynamics in a nonlinear fiber \cite{Solli}. Abscissae in (a) and (b) have been rescaled to equal units in terms of autocorrelation width, see insets. Solid lines: long-term averages. Dashed horizontal lines: significant wave height (SWH). Whenever applicable, the linear $1/\rm{e}$ autocorrelation width is shown as dashed vertical lines in the insets.\label{fig:example}}}
\vspace{-1em}
\end{figure}

Table \ref{tab:rogue} shows a comparison of the different parameters of the three data sets. We include the shape parameter $\beta$ of the histograms of the respective physical quantity $x$, as determined by a fit of the probability density function (PDF) to the Weibull function \cite{Weibull}
\begin{equation}
{\rm PDF}_{\rm W}(x) = \frac{\beta}{\alpha} \left(\frac{x}{\alpha}\right)^{\beta - 1}
\exp \left[ -\left(\frac{x}{\alpha} \right)^\beta \right].
\label{eq:weibull}
\end{equation}
Both optical system exhibit sub-exponential PDFs with minimum values of $\beta<1$, while the ocean records exhibit $\beta \approx 2$. All distributions are far away from normally distributed data, which would be indicated by $\beta > 10$. The dynamics in all systems exceed the SWH or an equivalently defined measure by a factor $>$2. In this linear statistical comparison of the data, there is another slightly more hidden aspect that sets the fiber data apart from the other systems. The temporal autocorrelation of the fiber system is essentially a delta function, whereas the other two time series exhibit autocorrelation widths $t_{\rm corr}$ that are clearly larger than the sampling interval $\delta t$. Figure~\ref{fig:example} shows characteristic rogue events for the three systems as well as their autocorrelation functions. To simplify inter-system comparison, time axes in Figs.~\ref{fig:example}(a,b) have been scaled in units of the autocorrelation width.

\begin{table}[tbh]
\begin{ruledtabular}
\begin{tabular}{l|ccccc}
Data set & $N$ & $\delta t$ & $\tau_{\rm corr}$ & $\beta$ & Excess SWH \\ \hline
Fiber & 5,000 & 40\,ns & $<$1\,ps & 0.87 & 4.2 \\
Filament & 60,000 & 1\,ms & 12\,ms & 0.44 & 10 \\
Draupner1 & 2,520 & 0.47\,s & 20\,s & 2.04 & 2.15  \\
Draupner2 & 2,520 & 0.47\,s & 20\,s & 1.8 & 2.3  \\
\end{tabular}
\end{ruledtabular}
\vspace{-1em}
\footnotesize{\caption{Intersystem comparison of rogue wave characteristics. $N$ : total number of samples in time series. $\delta t$ : sampling interval. $t_{\rm corr}$ : linear autocorrelation time ($1/e$ width). $\beta$ : Shape factor determined by Weibull fit to the histograms \cite{Weibull}. Excess SWH: Maximum wave height observed relative to significant wave height. Draupner1 was recorded on 01/01/1995 starting at 3:20pm; Draupner2 on 01/19/1995 at 11pm.\label{tab:rogue}}}
\vspace{-1em}
\end{table}

For a deeper analysis of the time series records, we employed one of the standard methods of nonlinear time series analysis, namely the Grassberger-Procaccia algorithm (GPA) \cite{Grassberger}. Let us suppose that a physical quantity $x$ has been sampled at a sampling interval $\delta t$ with a record length $N$. This yields a time series $\mathbf{x}$ that we denote by $\{x_1, x_2, x_3, \dots ,x_N\}$ in the following. From these time series, we select sub-series $\mathbf{y}_{i,m}=\{ x_i, x_{i+1}, \dots , x_{i+m} \}$. Here the record length $m$ is called the embedding dimension. These sub-series are then analyzed according to their Euclidian distances
\begin{equation}
r_{ijm}=\Big\|\mathbf{y}_{i,m}-\mathbf{y}_{j,m}\Big\|=\sqrt{\sum_{k=i}^{i+m} \left|x_k-x_{k+j-i}\right|^2},
\label{eq:euclid}
\end{equation}
which are accumulated for all $i$ and $j>i$ in a histogram
\begin{equation}
C_m(r)= \frac{ 2 \Big| \big\{r_{ijm} : (1-\epsilon)r<r_{ijm}\le r \big\}\Big|}{(N-m)(N-m-1)},
\end{equation}
where $\epsilon$ determines the bin size. Deterministic behavior is indicated by anomalously large counts $C_m(r)$ at $r=0.01$ to $0.1 r_{\rm max}$, where $r_{\rm max}$ denotes the largest non-zero bin of the histogram. In an illustrative way of speaking, such an anomaly can be understood as ``d\'ej\`a-vus'', i.e., the frequent appearance of certain time sequences $\mathbf{y}$ that are beyond a completely random distribution of events. Detection of such a statistical anomaly may be fairly obvious in physical systems with a low internal number of degrees of freedom, see for example \cite{Lorenz}. In turbulent systems, however, it requires very careful checks to judge the presence of determinism in a data set. As a null hypothesis, one therefore compiles surrogate data sets that resemble the original data set in all aspects of linear descriptive statistics. In particular, these surrogates feature an identical histogram $C_m(r)$ and also exhibit nearly the identical power spectrum and correlation function. We generated appropriate surrogate data in a three-step procedure, starting with the amplitude adjusted Fourier transform originally proposed by Theiler {\em et al.}~\cite{Theiler2}. To remove any residual bias at low Fourier frequencies, we applied an iterative method suggested by Schreiber and Schmitz \cite{Schreiber1}. Finally, we apply an annealing method to further improve agreement of the linear correlation function between original and surrogate \cite{Schreiber2}. All three steps reorder the time series without affecting the histogram. The randomization procedure heavily relies on the quality of true random number sequences that need to be free from any residual correlation. This constraint immediately rules out algorithmic random number generators. We therefore resorted to two approved sources of true random data \cite{randomorg,Marsaglia}, which have been generated by hardware generators. These random number generators are based on Johnson noise of a resistor or similar sources of uncorrelated noise. Additionally, both sources have been carefully checked to comply with a suite of statistical torture tests \cite{Marsaglia}.

\begin{figure}[tb]
\centering
\includegraphics[width=\linewidth]{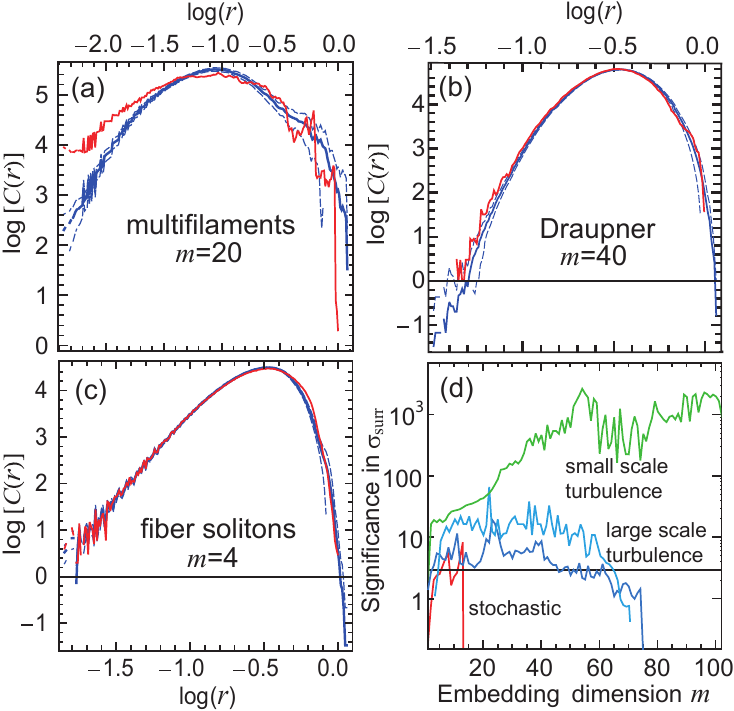}
\vspace{-1em}
\footnotesize{\caption{Nonlinear correlation analysis of rogue-wave systems. Shown is the histogram count $C(r)$ as a function of Euclidian distance $r$, cf.~Eq.~(\ref{eq:euclid}). Red line: histogram of original data set. Blue line: average of surrogates. Dashed blue lines: surrogate standard deviation spread. Embedding dimensions $m$ have been adapted to the correlation widths in (a) the multifilament system and (b) an analysis of the Draupner1 event. (c) Analysis for solitons in a nonlinear fiber, $m=4$ has been chosen to avoid trivial behavior. (d) Maximum detected statistical significance of deterministic behavior as a function of $m$. Symbols indicate the position of the histograms (a-c) in the plot. Significances $<3$ can safely be ignored as they indicate stochastic behavior within the expected credibility limits. Green curve: multifilament. Dark blue curve: Draupner1 data set. Light blue curve: Draupner2 data set. Red curve: fiber solitons. \label{fig:histo}}}
\vspace{-1em}
\end{figure}

Figure \ref{fig:histo} shows results of our surrogate analysis for the three different rogue wave systems. As optical rogue waves only appear in a fraction of the spectrum or beam profile, we have selected the respective time series that exhibit the most pronounced heavy tail, as indicated by the smallest value of the shape parameter $\beta$. Figure \ref{fig:histo} shows exemplary histograms $C_m(r)$ for the original data sets as well as for 100 surrogates. In these examples, the record length was adapted to the equivalent of one correlation length; for the fiber soliton system with its vanishing correlation length, we employed $m=4$. It needs to be emphasized that the below conclusions are reproduced for a large range of record lengths. Independent of $m$ and the system under consideration, there are no statistically significant deviations between original and surrogates at large $r$ as expected. Determinism in the data shows up as a lower histogram count $C_m(r)$ for the surrogates at low~$r$. This behavior appears most pronounced for the multifilament system [Fig.~\ref{fig:histo}(a)], with a statistical significance reaching values of $100\,\sigma_{\rm surr}$ well beyond the autocorrelation time of the system (equivalent to $m=12$). This finding further supports small-scale turbulence as the origin of the observed dynamics \cite{OurPRL,Milchberg,Dantus}. In contrast, the fiber optical system [Fig.~\ref{fig:histo}(c)] exhibits maximum statistical significances in the vicinity of $3\,\sigma_{\rm surr}$.  At larger values of $m$ and at low values of $r$, the surrogate histograms tend to lie even above the original histogram, which has to be interpreted as stochastic behavior, i.e., an absence of determinism in the data. It therefore seems to be evident that the fiber system is driven by a completely uncorrelated noise source. Considering the underlying physics of the system, amplified spontaneous emission noise of the pump laser \cite{Corwin,Dudley} appears the only possible candidate for such a noise source. The nonlinear fiber propagation itself is a completely deterministic process, which is ruled by the nonlinear Schr\"odinger equation. Nevertheless, this system is infamous for its large noise amplification \cite{Corwin,noiseamp}, which gives rise to a rapid loss of coherence during the highly nonlinear propagation. Analysis of the Draupner data sets [Fig.~\ref{fig:histo}(b)] is indicative of some determinism in the data. However, this determinism does not extend much beyond $\tau_{\rm corr}$ (equivalent to $m \approx 40$). Such a weak deterministic character of ocean surface waves has been reported before \cite{Elgar}, yet without including rogue waves in the analysis. This behavior is certainly different from the stochastic behavior in Fig.~\ref{fig:histo}(c). As an important conclusion, we find that the degree of determinism is completely independent of the extreme-value statistics, i.e., there is no correlation between the parameter $\beta$ and the predictability in the system. Moreover, there actually seem to be at least two different classes of rogue waves, those that emerge during turbulence and a second class that is generated by quantum noise.

\begin{figure}[tb]
\centering
\includegraphics[width=\linewidth]{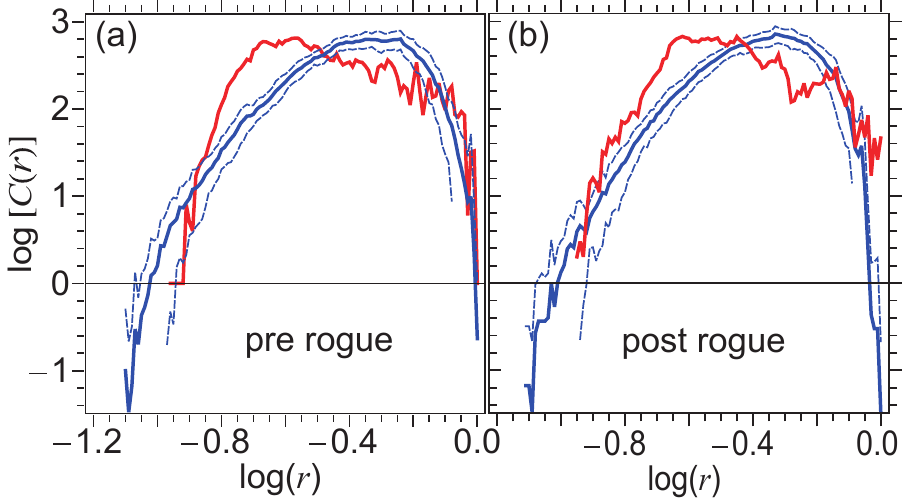}
\vspace{-1em}
\footnotesize{\caption{Histogram GPA analysis of time series segments in immediate vicinity of a rogue event in the multifilament data (red curve) in comparison to arbitrarily chosen segments of equal length (solid blue curve). (a) Pre-rogue statistics. Reference time series immediately preceding 289 isolated rogue events. (b) Post-rogue statistics. The statistical spread for arbitrarily chosen data (100 data sets) is indicated by the dashed blue curves. \label{fig:predict}}}
\vspace{-1em}
\end{figure}

So far, we have drawn conclusions from time series that only contained a relatively small fraction of rogue events. In principle, it is conceivable that these rogue events are generated by a second underlying mechanism in the system, which does not exhibit the same determinism as the rest of the trace. Given the very few records of real ocean rogue waves, this issue appears difficult to decide, despite some attempts towards a systematic collection of ocean rogue events \cite{proc}. In contrast, the large number of rogue events in our turbulent multifilament system provides a unique opportunity to tackle this question. To isolate the dynamics accompanying rogue events, we select sub-series $\mathbf{y}^{(\rm rogue)}_{i-m-1,m}$ and $\mathbf{y}^{(\rm rogue)}_{i+1,m}$ immediately preceding or following a rogue event, respectively. The index $i$ was chosen such that it either marks the beginning or the end of a single rogue event exceeding the SWH by a factor 2. We end up with 289 carefully chosen rogue events in our data, ensuring that they occurred more than one second and 260\,$\mu$m on the camera chip apart from each other. For the null hypothesis, we selected time series for randomly chosen $i$ at the same pixels of the camera. We then independently compiled histograms of the rogue-wave related sub-series and compared them to the histograms of the random data sets. The results are shown in Fig.~\ref{fig:predict}. Looking again at the low-$r$ part of the histograms at about $20\%$ of $r_{\rm max}$, we notice a statistically significant anomaly similar to the one in Fig.~\ref{fig:histo}, i.e., at certain values of $r$ we notice up to 5 times as many counts $C_m(r)$ for the rogue data sets than for arbitrarily chosen data. Nearly the same statistical anomaly is observed for the segments right after a rogue event [Fig.~\ref{fig:predict}(b)]. This means that the appearance of a rogue event is accompanied by characteristic dynamics in the time series. Similar peculiarities were observed in a wide range of fields, reaching from the appearance of epileptic seizures in medicine \cite{epileptic} to geomagnetic storms \cite{Chen}. The concomitant contraction of the parameter space can be taken as an analogy to the famous calm that precedes a storm. In nonlinear dynamics, such phenomena are typically understood as a shrinking of phase space connected to the appearance of the event, which, in turn, indicates the existence of at least one strange attractor. Most importantly, this finding appears to contrast the statement that ``rogue waves appear from nowhere and disappear without a trace''.

The observed predictability is rather robust, even if we select shorter sub-series at one or two correlation times distance to the rogue event. While the statistical significance certainly drops compared to Fig.~\ref{fig:predict}, we only start to lose evidence for predictability when we select segments more than $\approx 5 \, \tau_{\rm corr}$ away from the rogue event. One can certainly not directly transfer this finding to ocean rogue waves. At best one may expect to predict an ocean rogue wave a few ten seconds before impact, and it would require many future sightings to isolate characteristic patterns preceding an ocean rogue wave. Therefore any practical rogue wave prediction appears not overly realistic, despite the determinism in the system.

At first glance, the complete absence of determinism in one of the systems may actually be surprising as all three considered systems are described by variants of the Nonlinear Schr\"odinger Equation (NLSE) and exhibit similar soliton-like solutions. This formal similarity has always been considered a unifying aspect of rogue-wave supporting systems \cite{Roguereview}. The appearance of turbulence presupposes a spatially extended character of the system, and it is striking that both the ocean and the multifilament system are two-dimensional systems whereas the nonlinear fiber scenario is a one-dimensional system, which can also show turbulent behavior \cite{PhysicaD,oneD}. Nevertheless, the driver for rogue waves in the fiber system is quantum noise of the supercontinuum pump laser, giving rise to a complete lack of determinism in the observed dynamics. In some sense of speaking, therefore, extreme events in the fiber optic context appear as the better rogue waves because of their guaranteed unpredictability. In the original ocean context, in principle, rogue waves seem to be predictable on short time scales, which further demystifies them as a deterministic phenomenon.

We gratefully acknowledge Daniel Solli (UCLA, Los Angeles, California) for providing original rogue wave data for the fiber-optical system and Statoil ASA for the permission to use the data recorded on the Draupner oil platform in January 1995. Mads Haahr is acknowledged for giving us generous access to true random number series at random.org. Moreover, we thank Michel Olagnon (IFREMER, Brest, France), Sverre Haver (Norwegian University of Science and Technology, Trondheim), and Anne Karin Magnusson (Norwegian Meteorological Institute, Oslo) for their support. We further acknowledge Fedor Mitschke (University of Rostock) and Claus Ropers (University of G\"ottingen) for helpful discussions.

\footnotesize{
}

\end{document}